\documentclass[intlimits,twoside,a4paper]{article}

\usepackage{graphicx}        

\usepackage[eqsecnum]{cmpj3}

\issue{2026}{29}{2}{23802}
\doinumber{10.5488/CMP.29.23802}

\title[Binary hard-disk mixture on a patterned adhesive surface]%
{Monte Carlo simulation of selective adsorption in a binary hard-disk mixture on patterned adhesive surfaces %
}
\author[N.~Kukarkin, T.~Patsahan]{N.~Kukarkin\orcid{0009-0005-8787-5639}\refaddr{label1},
T.~Patsahan\orcid{0000-0002-7870-2219}\refaddr{label1,label2}\thanks{Corresponding author: \email{tarpa@icmp.lviv.ua}.}}
\addresses{
\addr{label1} Yukhnovskii Institute for Condensed Matter Physics of the National Academy of Sciences of Ukraine, 1~Svientsitskii~Str., 79011 Lviv, Ukraine
\addr{label2} Institute of Applied Mathematics and Fundamental Sciences, Lviv Polytechnic National University, 12~S.~Bandera~Str., 79013 Lviv, Ukraine}
\date{Received 19 May 2026; revised 3 June 2026; accepted 3 June 2026; published 29 June 2026}

\begin{document}

\maketitle

\begin{abstract}
Selective adsorption in a two-dimensional model of a binary hard-disk mixture on
patterned adhesive surfaces is studied using grand canonical Monte Carlo
simulations. The two species have equal diameters and equal bulk chemical
potentials, but different attraction strengths to adhesive domains. Thus,
affinity-driven selectivity is separated from particle-size asymmetry and unequal
chemical potentials. The surface pattern is defined by domain size, domain surface
coverage, and ordered or disordered arrangement of circular domains. The results
show that selectivity strongly depends on surface geometry, especially at low and
intermediate chemical potentials. Domains comparable to the particle size enhance
selectivity by forming adsorption regions with large particle--domain overlap,
whereas larger domains can provide high selectivity at low chemical potentials.
For small domains, further reduction in size can also increase selectivity as the
system approaches a uniform attractive surface with corresponding effective
affinity parameters of the species.
\end{abstract}

\section{Introduction}

Adsorption of micro- and nanoscale particles on chemically patterned surfaces is a
fundamental problem in interfacial science, soft matter physics, and materials
engineering. Patterned substrates provide a means of controlling both the total amount of
adsorbed particles and their spatial organization at the surface.
This is important for the design of functional coatings, colloidal assembly templates,
biosensors, protein and nanoparticle arrays, and platforms for controlled attachment
of biological objects such as cells~\cite{Maury2008,VanDommelen2018, Burkhardt2010,Xing2020,Jambhulkar2024,Rao2024}. 
In such systems, adsorption is determined by an interplay between energetic and geometric factors: particles are
attracted to favourable surface regions, while their lateral arrangement is constrained
by excluded volume interactions and by the spatial distribution of adhesive sites.

The influence of surface heterogeneity has been studied in a variety of model and
experimental systems. In polymer adsorption, Monte Carlo simulations have shown
that the size of chemically distinct surface domains can strongly affect adsorption
and ordering. In particular, efficient adsorption may occur when the domain size is
comparable to the characteristic size of the adsorbing object or its structural
units~\cite{SemlerGenzer2003,SemlerGenzer2004}. Related theoretical work has
shown that stochastic surface heterogeneity can also enhance adsorption by creating
a distribution of favourable local environments~\cite{ChervanyovHeinrich2006}.
Simple lattice and continuum models of adsorption on heterogeneous substrates
further demonstrate that the concentration, spatial arrangement, and geometry of
active sites can strongly influence adsorption isotherms, local ordering, and surface
coverage~\cite{Nitta1997,TalbotTarjusViot2008,OleyarTalbot2007}.

Irreversible particle deposition and random sequential adsorption on patterned or
heterogeneous surfaces represent another important class of related problems. In these
models, the adsorbed particles usually remain fixed after deposition, so that the
final structure is controlled by surface blocking and packing constraints. Such
studies have shown that substrate patterning can significantly modify the approach
to the jamming limit, the morphology of deposited layers, and the crossover between
lattice-like and continuum-like behaviour~\cite{CadilheAraujoPrivman2007,AraujoCadilhePrivman2008,Marques2012,PrivmanYan2016,StojiljkovicVrhovac2017}.
Although these models describe irreversible rather than equilibrium adsorption,
they provide important physical insight into the role of surface geometry and
excluded volume effects in the formation of adsorbed layers.

Multicomponent adsorption on heterogeneous surfaces is of particular interest
because, in many practical situations, different adsorbates compete for the same
surface regions. In such systems, the adsorption behaviour is determined not only
by the properties of the individual components, but also by the way the available
adsorption sites are distributed between them. Adsorption of binary mixtures on heterogeneous surfaces was investigated
using a range of theoretical and simulation approaches, including lattice models,
integral equation methods, cluster approximations, and Monte Carlo
simulations~\cite{Nieszporek2005,SanchezVarretti2014,SanchezVarretti2019}.
These studies showed that surface heterogeneity, lateral interactions, and
differences in adsorption energy between species can strongly affect adsorption isotherms,
the composition of the adsorbed layer, and surface ordering.

Binary and multicomponent particle mixtures have also been studied in the context
of irreversible deposition. Random sequential adsorption of two component mixtures
of disk particles was used to analyze how the particle size ratio and relative
adsorption rates affect the jamming coverage and the composition of the deposited
layer~\cite{Wagaskar2020}. Competitive adsorption of particles with different
shapes, such as disks and discorectangles, has showed that mixtures may exhibit
packing and percolation behaviour that cannot be inferred from the corresponding
one component systems~\cite{Lebovka2024}. Other studies of binary hard core
particles on lattices addressed the relation between adsorption kinetics,
surface diffusion, and thermodynamic properties~\cite{Darjani2021}. These works
highlight the general importance of competition between different adsorbing species,
but they are often focused on irreversible deposition, lattice models, or mixtures
with different particle sizes or shapes.

Selective adsorption is especially important when different particles have different
affinities to a structured surface. This situation is common in systems involving
functionalized nanoparticles, polymer brush surfaces, affinity sensors, proteins,
and cells~\cite{Steinbach2013,Beggiato2022,KumarParajuliHahm2007, Didar2010adhesion,Badenhorst2025}. 
For example, patterned or chemically structured surfaces can be used to enhance adsorption efficiency, to guide the spatial
organization of nanoparticles or biomolecules, and to promote preferential
attachment of selected biological objects. However, in real systems, selectivity is
usually affected by several factors at once, including particle size, shape, charge,
deformability, specific binding, and kinetic effects. Therefore, minimal statistical
mechanical models are useful for separating the energetic and geometric contributions to selective adsorption.

In our previous study, equilibrium adsorption of a one component system of hard disks
on patterned adhesive surfaces was investigated using Monte Carlo simulations~\cite{KukarkinPatsahan2026}. 
The adhesive domains were arranged either on a square lattice or in a disordered
configuration, and the attraction between a particle and a domain was assumed to be
proportional to their overlap area. It was shown that
adsorption is controlled not only by the total surface coverage of adhesive domains,
but also by their size and spatial arrangement. The most pronounced geometric
effect was observed when the domain size was comparable to the particle size. In
this case, favourable locations in which a particle strongly overlaps with an
individual domain noticeably affected the adsorption isotherms and the
structure of the adsorbed layer. At high particle coverages, however, the influence
of the surface pattern became weaker because excluded volume and particle packing
started to dominate~\cite{KukarkinPatsahan2026}.

The next step is to extend this model to mixtures. In the present work, we
consider a binary mixture of hard disks adsorbed on a patterned adhesive surface.
We assume that the two species of particles have the same diameter and the same chemical potential in the
bulk. Thus, they are identical with respect to hard-core interactions 
and have the same probability of exchange with the bulk reservoir. 
They differ only in the strength of their attractive
interaction with the adhesive domains. This formulation isolates affinity-driven
selectivity from other effects. Since the bulk mixture does not favour either component, the
symmetry between the two species is broken only at the patterned surface. Therefore, the
composition of the adsorbed layer  provides a direct measure of selective
adsorption caused by different particle-domain affinities. Using grand canonical
Monte Carlo simulations, we analyze the adsorption isotherms of both components and 
the selectivity coefficient. Thus, the model  provides a simple equilibrium approach for
examining how surface geometry, particle packing, and adsorption affinity jointly
control selective adsorption on patterned adhesive surfaces.

\section{Model description}

\begin{figure}[!htb]
	\centering
	\includegraphics[width=0.55\linewidth]{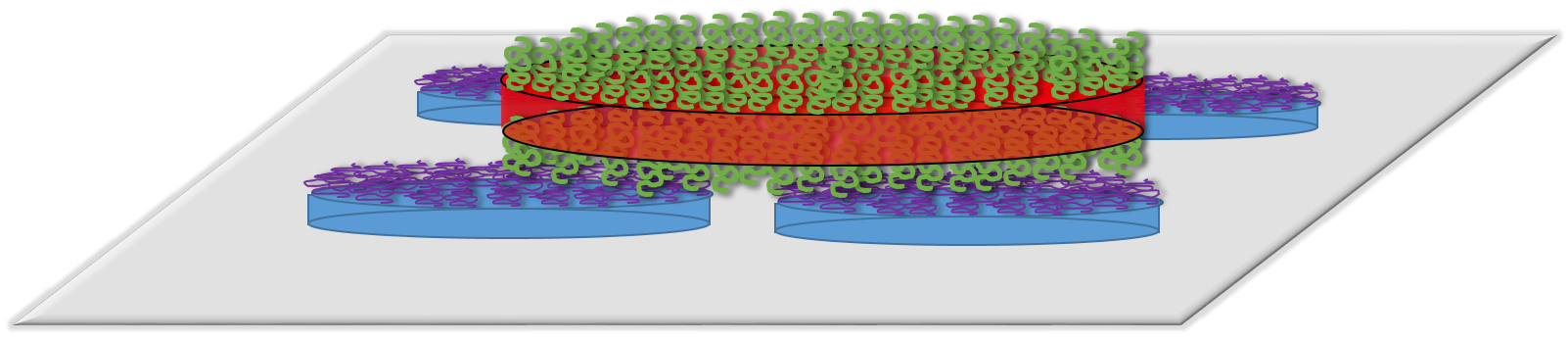}
        \caption{(Colour online) 
        Schematic representation of a functionalized disk-like particle that adsorbs on a
        surface patterned with circular adhesive domains. The red disk denotes the
        particle, while the blue {disks} represent adhesive domains uniformly covered with
        functional groups (violet) that interact specifically with groups on the particle
        surface (green).
        }
    \label{fig:sketch}
\end{figure}

The model developed in this work is inspired by the three-dimensional model of a
two-component system of particles that adsorb on a surface patterned with adhesive
domains introduced in~\cite{Badenhorst2025} in order to study cell sorting
phenomena. Here, we use the two-dimensional version of this model, recently
adapted in~\cite{KukarkinPatsahan2026} for a one-component system, and extend
it to a binary hard-disk mixture. {The two-dimensional description corresponds to the  adsorption of a single monolayer of particles on the patterned surface, which  represents the first adsorbed layer in the corresponding three-dimensional system}.
The system consists of hard disks moving in the plane of a surface patterned with
circular adhesive domains. The number of adsorbed particles is controlled by the chemical potential of the
bulk reservoir, which is in thermodynamic equilibrium with the adsorbed layer.
The domains are arranged either on a regular square lattice or in a disordered configuration. 
The particle--domain interaction is attractive and is taken to be proportional to
the overlap area between a particle and an adhesive region. This describes adhesion arising from contact between 
the lower surface of the particle and the tops of the domains, which are uniformly functionalized
with chemical groups that interact specifically with groups located on the particle
surface, as schematically shown in figure~\ref{fig:sketch}.

In a binary mixture, the strength of this interaction depends on the particle
type 
We follow the
concept proposed in~\cite{KukarkinPatsahan2026} and use the same form of the
particle--domain interaction potential, but write it in a more general form by
introducing the particle species index $\alpha$. Thus, the adsorption energy of
particle $i$ of species $\alpha$, located at the surface in $(x_i,y_i)$ coordinates, is defined as the
sum of pair interaction energies with all adhesive domains overlapping with this
particle:
\begin{equation}
	U_{\alpha\mathrm{s}}^{\mathrm{ads}}(x_i,y_i)
	=
	\sum_{\{j \, | \, r_{ij} \leqslant r_{\alpha\mathrm{d}}\}}
	A_{\alpha\mathrm{d}}\, s_{\alpha\mathrm{d}}(r_{ij}),
	\label{eq:Up_ads}
\end{equation}
where $r_{\alpha\mathrm{d}}=R_{\alpha}+R_{\mathrm d}$
is the interaction range between a particle of species $\alpha$ and a domain,
and $A_{\alpha\mathrm{d}}$ is the attraction strength between them, here the subscript `$\mathrm{d}$' denotes the adhesive domain
{and the subscript `$\mathrm{s}$' means `surface'}.
The sum in equation~\eqref{eq:Up_ads} runs over all domains $j$ with $(x_j,y_j)$ coordinates and whose centers are located within
the distance $r_{\alpha\mathrm{d}}$ from particle $i$. This center-to-center
distance is given by $r_{ij}=\sqrt{(x_i-x_j)^2+(y_i-y_j)^2}$.
The overlap area between a particle $i$ of radius $R_{\alpha}$ and a domain $j$ of radius $R_{\mathrm d}$ is
\begin{equation}
	s_{\alpha\mathrm{d}}(r_{ij})=
	\begin{cases}
		\min(s_{\alpha},s_{\mathrm d}), 
		& r_{ij}\leqslant |R_{\alpha}-R_{\mathrm d}|,\\[6pt]
		s_{\mathrm{int}}(r_{ij}), 
		& |R_{\alpha}-R_{\mathrm d}|< r_{ij}\leqslant R_{\alpha}+R_{\mathrm d},\\[6pt]
		0, 
		& r_{ij}>R_{\alpha}+R_{\mathrm d},
	\end{cases}
	\label{eq:overlap_piecewise}
\end{equation}
where $s_{\alpha} = \piup R^2_{\alpha}$ and $s_{\mathrm d}=\piup R^2_{\rm{d}}$ are the areas of a particle of species
$\alpha$ and of a domain, respectively. The partial overlap area $s_{\mathrm{int}}(r_{ij})$ is calculated using the
standard formula for the intersection area of two circles whose centers are
separated by a distance $r_{ij}$:
\begin{equation}
	\begin{aligned}
		s_{\mathrm{int}}(r_{ij}) &=
		R_{\alpha}^{2}
		\arccos\!\left(
		\frac{r_{ij}^{2}+R_{\alpha}^{2}-R_{\mathrm d}^{2}}
		{2R_{\alpha}r_{ij}}
		\right)
		+
		R_{\mathrm d}^{2}
		\arccos\!\left(
		\frac{r_{ij}^{2}+R_{\mathrm d}^{2}-R_{\alpha}^{2}}
		{2R_{\mathrm d}r_{ij}}
		\right)\\
		&\quad
		-\frac{1}{2}
		\sqrt{
			\left[(R_{\alpha}+R_{\mathrm d})^2-r_{ij}^2\right]
			\left[r_{ij}^2-(R_{\mathrm d}-R_{\alpha})^2\right]
		}.
	\end{aligned}
	\label{eq:intersection_area}
\end{equation}
Particles can move freely in the plane of the surface and interact with the adhesive domains through the potential $U_{\alpha\mathrm{s}}^{\mathrm{ads}}$. Particle--particle interactions are described by hard-disk repulsion. 
Adsorption and desorption of particles at the surface are treated within the grand
canonical ensemble, where the number of adsorbed particles of species $\alpha$ is
controlled by the corresponding chemical potential $\mu_\alpha$ of the bulk
reservoir. {Lower values of the  chemical potential correspond to 
lower bulk particle density, whereas higher values correspond to 
denser bulk conditions and thus promote stronger adsorption.}
For the binary mixture of species `1' and `2', we further restrict ourselves to the symmetric case $\mu_1=\mu_2=\mu$.

It is worth noting that the simplified model presented above focuses on equilibrium adsorption and the formation of
a particle monolayer, without considering the details of bulk behaviour. 
Instead, the model is designed to capture the main effects arising from particle-domain
interactions, excluded volume, and competition between particles for the
most favourable adsorption positions. These effects are controlled by the domain
surface coverage, the lateral arrangement of the domains, either ordered or
disordered, and the domain size. 
Although more general cases are possible, in the present study we consider
the particles of equal size. This choice allows us to isolate the effects directly associated with the geometry of the patterned surface 
and to examine how this geometry influences the selective adsorption driven solely by differences in particle-domain affinity. 
Following this idea we raise the questions how the size of domains affects the selective adsorption in a binary
mixture where the two components differ only in their affinity for the adhesive domains, and how the geometry of the surface pattern
modifies this effect.

\section{Computer simulation details}

To address the questions posed above, we perform grand canonical Monte Carlo (GCMC)~\cite{allen2017computer}
simulations of a binary mixture of hard disks, with species labelled by
$\alpha=1$ and $2$, in a square simulation box of linear size $L$. The particles
can move in two dimensions on a patterned surface containing circular adhesive domains arranged according
to the specified surface geometry. The GCMC simulations allows us to obtain the adsorption isotherms of both particle
species, $\sigma_1(\mu^*)$ and $\sigma_2(\mu^*)$. Here, $\sigma_\alpha=\piup R_\alpha^2\rho_\alpha$ is the surface coverage by particles of
species $\alpha$, and $\rho_\alpha=N_\alpha/L^2$ is the corresponding surface number density. 
The reduced chemical potential is defined as
$\mu^*=\mu/k_{\rm B}T$, where $k_{\rm B}$ is the Boltzmann constant and $T$ is
the temperature. In the simulations considered in the present study, the same
chemical potential is used for both species, $\mu^*_1=\mu^*_2=\mu^*$. Based on the
adsorption isotherms, we calculate the selectivity coefficient
$S_{1/2}(\mu^*)=\sigma_1(\mu^*)/\sigma_2(\mu^*)$, which quantifies the
preferential adsorption of species `1' relative to species `2'. 
We assume that particles of species `1' interact with the adhesive domains twice as
strongly as particles of species `2'. The corresponding attraction parameters are
chosen as
$A_{1\mathrm d}=\beta A_0/(\piup R_{\rm p}^2)=-12.7324 D_{\rm p}^{-2}$ and
$A_{2\mathrm d}=A_{1\mathrm d}/2=-6.3662 D_{\rm p}^{-2}$, where
$\beta=1/k_{\rm B}T$ and $|A_0|=10k_{\rm B}T$, indicating that the magnitude of the full-overlap adhesion energy 
of a particle of species `1' is ten times larger than the thermal energy, while $D_{\rm p}=2 R_{\rm p}$ is the diameter
of particles.
This choice ensures reliable adsorption of both species even when particles only partially overlap with
adhesive domains. Since species `2' has a weaker attraction to the domains than species `1', it is expected to adsorb
less strongly than species `1', which should be directly reflected in the resulting adsorption selectivity $S_{1/2}(\mu^*)$. 
Our aim is to determine how the geometry of the surface pattern affects the
magnitude of this selectivity and under which conditions it becomes most
pronounced. 

GCMC simulations were performed for a binary mixture of hard disks of species
`1' and `2' adsorbed on patterned surfaces with adhesive domains of
different sizes.
Three domain sizes were considered, i.e., $D_{\rm d}/D_{\rm p}=0.5$, $1.0$, and $2.0$, where
$D_{\rm d}=2R_{\rm d}$ is the diameter of a domain. 
{The particles of both species are of the same size}, $D_1=D_2=D_{\rm p}$. Two values of the domain surface coverage were considered,
$\sigma_{\rm d}=0.349$ and $0.503$, where $\sigma_{\rm d}=\piup R_{\rm d}^2 \rho_{\rm d}$,
and $\rho_{\rm d}=N_{\rm d}/L^2$ was the number density of adhesive domains. The
linear size of the simulation box was set to $L=90.0D_{\rm p}$ for $\sigma_{\rm d}=0.349$ and to $L=75.0D_{\rm p}$ for $\sigma_{\rm d}=0.503$.
Then, the number of domains was determined as $N_{\rm d}=\sigma_{\rm d} L^2/(\piup R_{\rm d}^2)$.
For the specified parameters, ordered domain patterns were generated as square
lattices containing $N_{\rm d}=N\times N$ adhesive domains. The number of lattice
nodes was set to $N=90$, $60$, and $30$ for
$D_{\rm d}/D_{\rm p}=0.5$, $1.0$, and $2.0$, respectively, for both values of
the domain surface coverage. For the disordered domain configurations, the same
numbers of domains were used at the corresponding values of $\sigma_{\rm d}$ and
$D_{\rm d}$. 
The initial configurations of particles were generated randomly, avoiding overlaps
between hard disks. For each set of parameters, the simulations were started from
an equimolar mixture containing $2000$ particles of each species in the simulation
box of size $L=90.0D_{\rm p}$ and $1500$ particles of each species in the box of
size $L=75.0D_{\rm p}$. 

During the GCMC simulation runs, three types of trial moves were used: particle
displacements, insertions, and deletions. Displacement moves sampled the positions
of particles already present at the surface, while insertion and deletion moves
changed the numbers of particles of each species, $N_1$ and $N_2$, according to the Metropolis criterion using the change in the potential
energy and the given chemical potential~\cite{allen2017computer}. 
The maximum distance of particle displacement for one move was set to $0.5 D_{\rm{p}}$. 
The fractions of the displacement, insertion and deletion trial moves was taken as $0.5$, $0.25$ and $0.25$, respectively.
Each simulation step corresponded to $N_1+N_2$ trial moves.
The system was equilibrated until the average numbers of particles of species `1' and `2' reached equilibrium values corresponding to the chemical
potential $\mu^*$.
The total number of steps in a simulation run was varied up to $2 \cdot 10^6$, depending
on the value of the reduced chemical potential. This was sufficient to equilibrate
the system and to obtain satisfactory statistics, including at the highest values
of $\mu^*$ considered. The reduced chemical potential was changed over the range
$-12.0 \leqslant \mu^* \leqslant 5.0$.
The values of average $N_1$ and $N_2$ were used to calculate the particle surface coverage for each species $\sigma_1$ and $\sigma_2$.

\begin{figure}[!htb]
	\centering
    \includegraphics[width=0.9\linewidth]{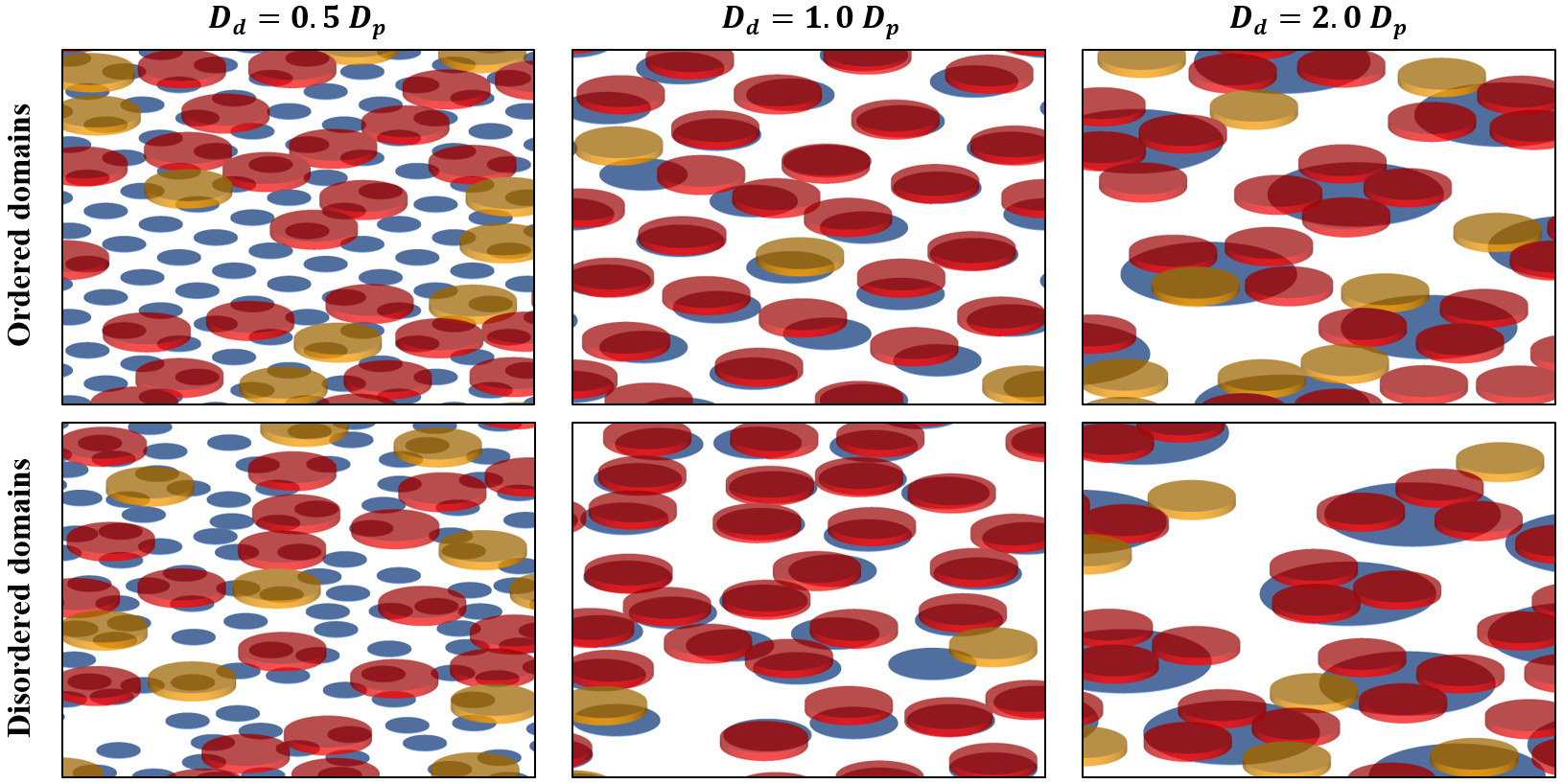}
	\caption{(Colour online) 
		Representative fragments of snapshots for a two-dimensional binary hard-disk
		mixture adsorbed on a patterned adhesive surface with domain surface coverage
		$\sigma_{\rm d}=0.349$ at the reduced chemical potential $\mu^*=-2.0$.
		Blue regions denote adhesive domains, while red and yellow disks denote particles
		of species 1 and 2, respectively. The adhesive domains are arranged on a square
		lattice in the top row and in a disordered configuration in the bottom row.
		From left to right, the panels correspond to domain sizes
		$D_{\rm d}/D_{\rm p}=0.5$, $1.0$, and $2.0$, respectively. Particle transparency
		is used to make the overlaps between particles and domains visible.
	}
	\label{fig:snapshots}
\end{figure}

Figure~\ref{fig:snapshots} shows representative fragments of simulation snapshots
for a binary mixture of hard disks adsorbed on patterned adhesive surfaces at the
reduced chemical potential $\mu^*=-2.0$. The snapshots illustrate both ordered
and disordered arrangements of circular adhesive domains at the same domain
surface coverage, $\sigma_{\rm d}=0.349$. The top row corresponds to domains
arranged on a square lattice, whereas the bottom row shows disordered domain
configurations. From left to right, the panels represent an increasing domain size,
$D_{\rm d}/D_{\rm p}=0.5$, $1.0$, and $2.0$, respectively. Blue regions denote
adhesive domains, while red and yellow disks correspond to particles of species `1'
and `2'. Since species `1' has a stronger attraction to the adhesive domains, the
snapshots provide a direct visual illustration of the competition between the two
components for favourable adsorption positions and of how this competition is
affected by the size and spatial arrangement of the domains.

\section{Results and discussion}

\begin{figure}[!htb]
	\centering
	\includegraphics[width=0.47\linewidth]{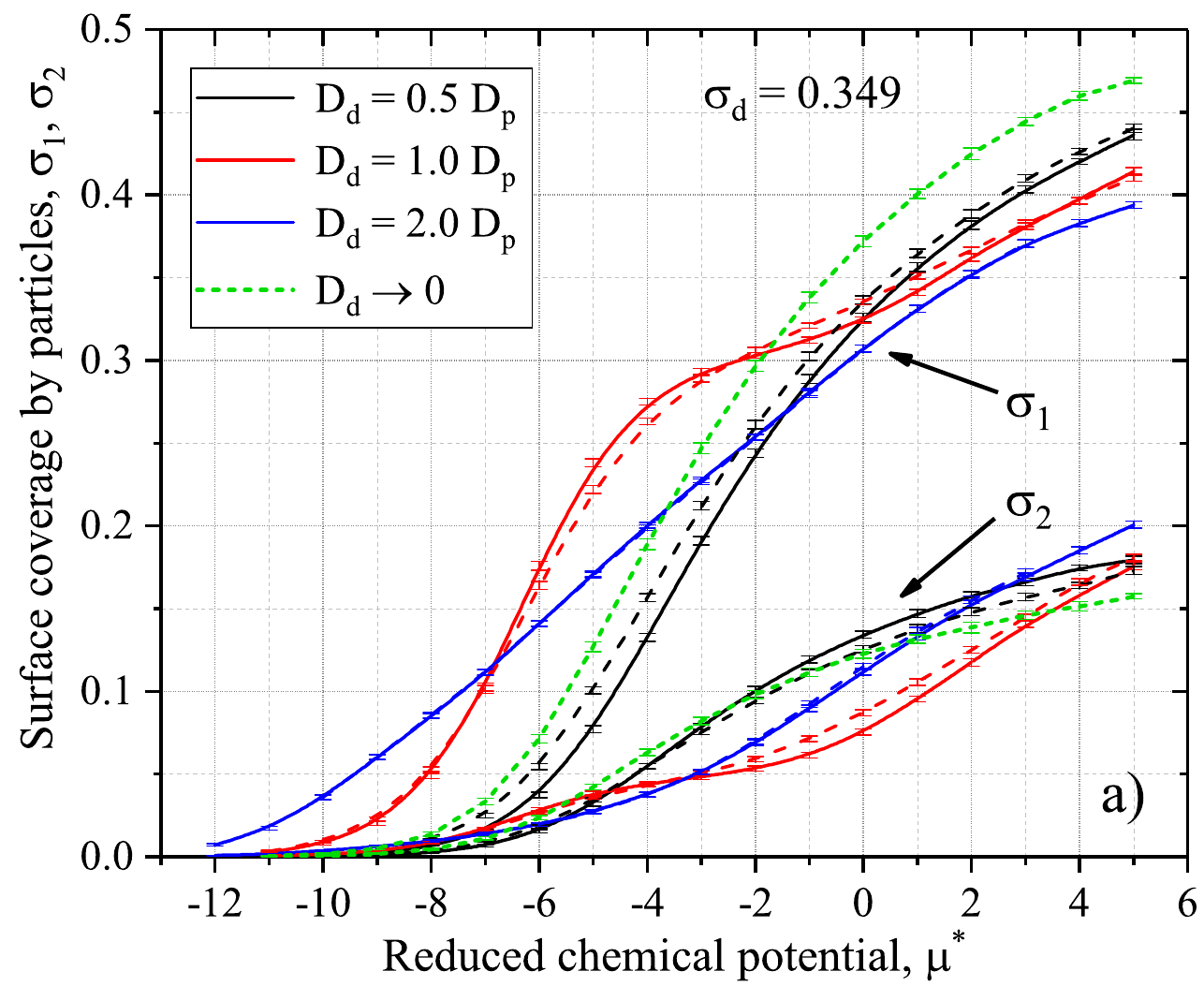}
	\includegraphics[width=0.47\linewidth]{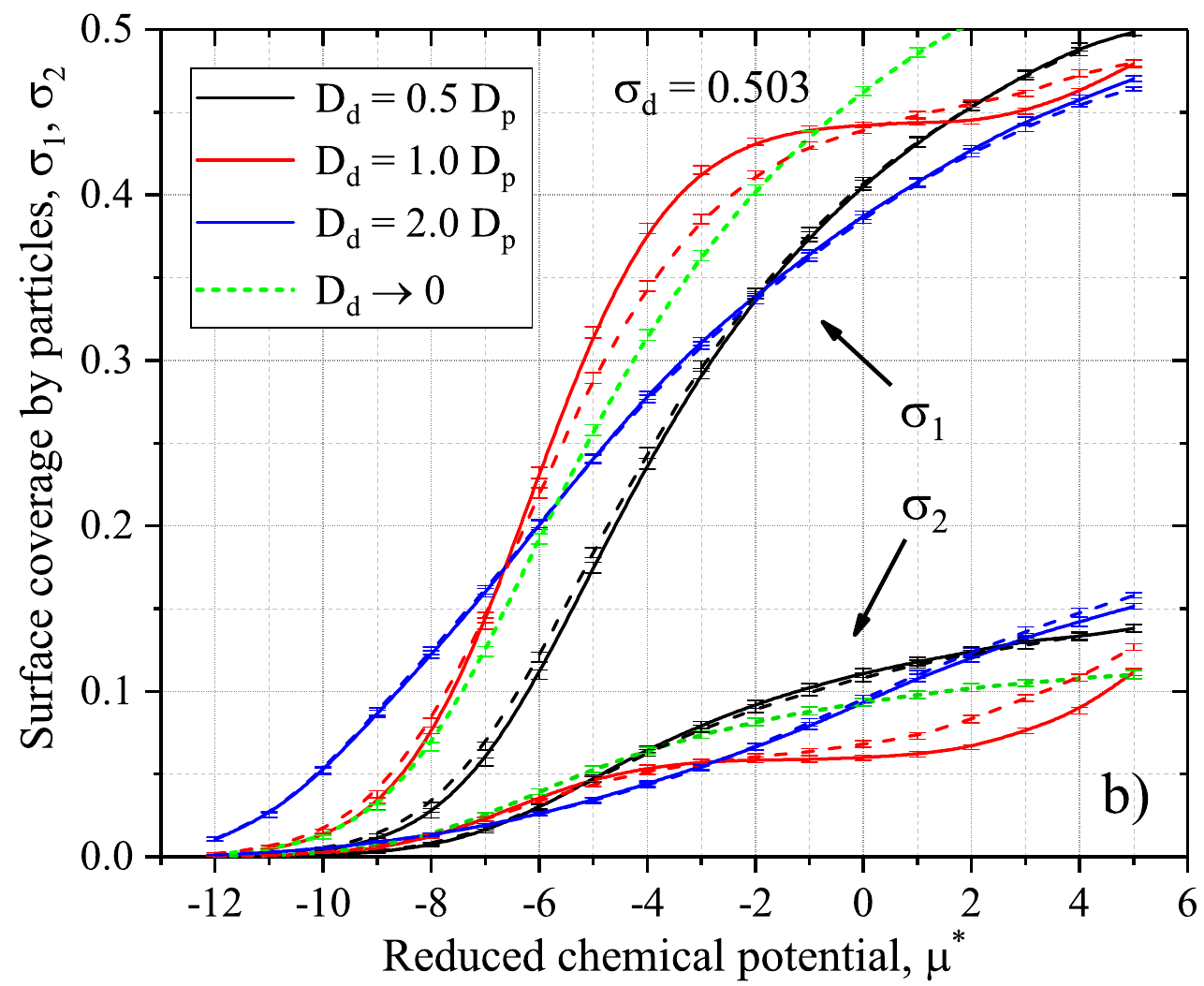}	
	\caption{(Colour online)
		Adsorption isotherms of a binary mixture of disk-like particles on a patterned
		adhesive surface with domain surface coverage
		$\sigma_{\rm d}=0.349$~(panel~a) and $\sigma_{\rm d}=0.503$~(panel~b).
		Black, red, and blue lines correspond to domain sizes
		$D_{\rm d}/D_{\rm p}=0.5$, $1.0$, and $2.0$, respectively.
		Solid lines denote ordered domain arrangements, while dashed lines denote disordered domain arrangements.
		Green short-dashed lines correspond to the limiting case of a uniform attractive
		surface, $D_{\rm d}\to 0$, with effective attraction parameters
		$A_{1\mathrm{d}}^{\rm eff}=-4.4444 D_{\rm p}^{-2}$ and
		$A_{2\mathrm{d}}^{\rm eff}=A_{1\mathrm{d}}^{\rm eff}/2$ in panel~a, and
		$A_{1\mathrm{d}}^{\rm eff}=-6.4 D_{\rm p}^{-2}$ and
		$A_{2\mathrm{d}}^{\rm eff}=A_{1\mathrm{d}}^{\rm eff}/2$ in panel~b.
	}	
	\label{fig:isotherms}
\end{figure}

{A series of grand canonical Monte Carlo simulations was performed for a binary
hard-disk mixture adsorbed on a surface patterned with monodisperse adhesive domains of diameter $D_{\rm d}$, 
which was varied at a fixed domain surface coverage $\sigma_{\rm d}$.
Two values of the domain surface coverages were considered in our study $\sigma_{\rm d}=0.349$ and $0.503$.
The adsorption isotherms obtained for both particle species are shown in
figure~\ref{fig:isotherms} as the surface coverages
$\sigma_1$ and $\sigma_2$ plotted against the reduced chemical potential $\mu^*$.
An increase in $\mu^*$ corresponds to more favourable particle insertion from the
bulk reservoir and therefore results in a higher number of particles adsorbed at
the surface.}
The upper set of curves corresponds to species `1', which has the stronger attraction to the adhesive domains, 
whereas the lower set corresponds to species `2'. As expected, $\sigma_1$ is larger than $\sigma_2$
over the whole range of chemical potentials for all considered surface patterns.
Thus, even though the two species have the same size and the same chemical
potential in the reservoir, the difference in particle--domain affinity leads to an
enrichment of the adsorbed layer by particles of species `1'.
The shape of the isotherms noticeably depends on the domain size. At
$\sigma_{\rm d}=0.349$ (figure~\ref{fig:isotherms}a), the adsorption of species `1'
is enhanced at low and intermediate chemical potentials when the domain size is
comparable to the particle size, $D_{\rm d}/D_{\rm p}=1.0$. In this case, particles
can occupy favourable positions with a large overlap with individual adhesive
domains, which leads to a faster increase of $\sigma_1$. For larger domains,
$D_{\rm d}/D_{\rm p}=2.0$, adsorption starts already at lower values of $\mu^*$,
but the growth of the isotherm is more gradual. For smaller domains,
$D_{\rm d}/D_{\rm p}=0.5$, the attractive regions are more numerous and more
uniformly distributed over the surface, so that the corresponding isotherm is
smoother and approaches the behaviour expected for an effective
uniform attractive surface.

Increasing the domain surface coverage to $\sigma_{\rm d}=0.503$
(figure~\ref{fig:isotherms}b) strengthens the effect of the patterned surface.
The adsorption of species `1' becomes more pronounced and shifts to lower chemical
potentials. This effect is especially visible for $D_{\rm d}/D_{\rm p}=1.0$, where
the isotherm of species `1' rises steeply and reaches high surface coverage already
at intermediate values of $\mu^*$. At the same time, the adsorption of species `2' is
strongly suppressed for this domain size. This indicates that the most favourable
adsorption positions are preferentially occupied by the more strongly attracted
particles of species `1', leaving fewer favourable positions available for species `2'.

The comparison with the limiting case $D_{\rm d}\to 0$ illustrates the trend
expected upon further decreasing the domain size and provides an insight into the
general effect of surface patterning. In this limit, the surface behaves as a
uniformly attractive substrate with an effective interaction strength determined
by the domain surface coverage. For the uniform surface equivalent to the patterned
surface with $\sigma_{\rm d}=0.349$, the effective attraction parameters are
$A_{1\mathrm{d}}^{\rm eff}=-4.4444 D_{\rm p}^{-2}$ and
$A_{2\mathrm{d}}^{\rm eff}=A_{1\mathrm{d}}^{\rm eff}/2$, while for $\sigma_{\rm d}=0.503$ they are
$A_{1\mathrm{d}}^{\rm eff}=-6.4 D_{\rm p}^{-2}$ and
$A_{2\mathrm{d}}^{\rm eff}=A_{1\mathrm{d}}^{\rm eff}/2$. As seen in
figure~\ref{fig:isotherms}, the corresponding isotherms for the uniform surface are
smoother than those obtained for finite domains. The deviations from this
reference case demonstrate that the adsorption behaviour is not determined solely
by the average attraction strength, but is also affected by the finite size and
spatial arrangement of the adhesive domains.

For $\sigma_{\rm d}=0.349$, the comparison between ordered and disordered domain
arrangements shows that positional disorder modifies the isotherms mainly at low
and intermediate chemical potentials. This is the regime where adsorption is most
sensitive to the local distribution of favourable adsorption sites. At higher
$\mu^*$, the difference between ordered and disordered patterns becomes less
pronounced, because the adsorbed layer becomes denser and its behaviour is
increasingly governed by excluded volume effects rather than by the details of the
surface pattern.

\begin{figure}[!htb]
	\centering
	\includegraphics[width=0.47\linewidth]{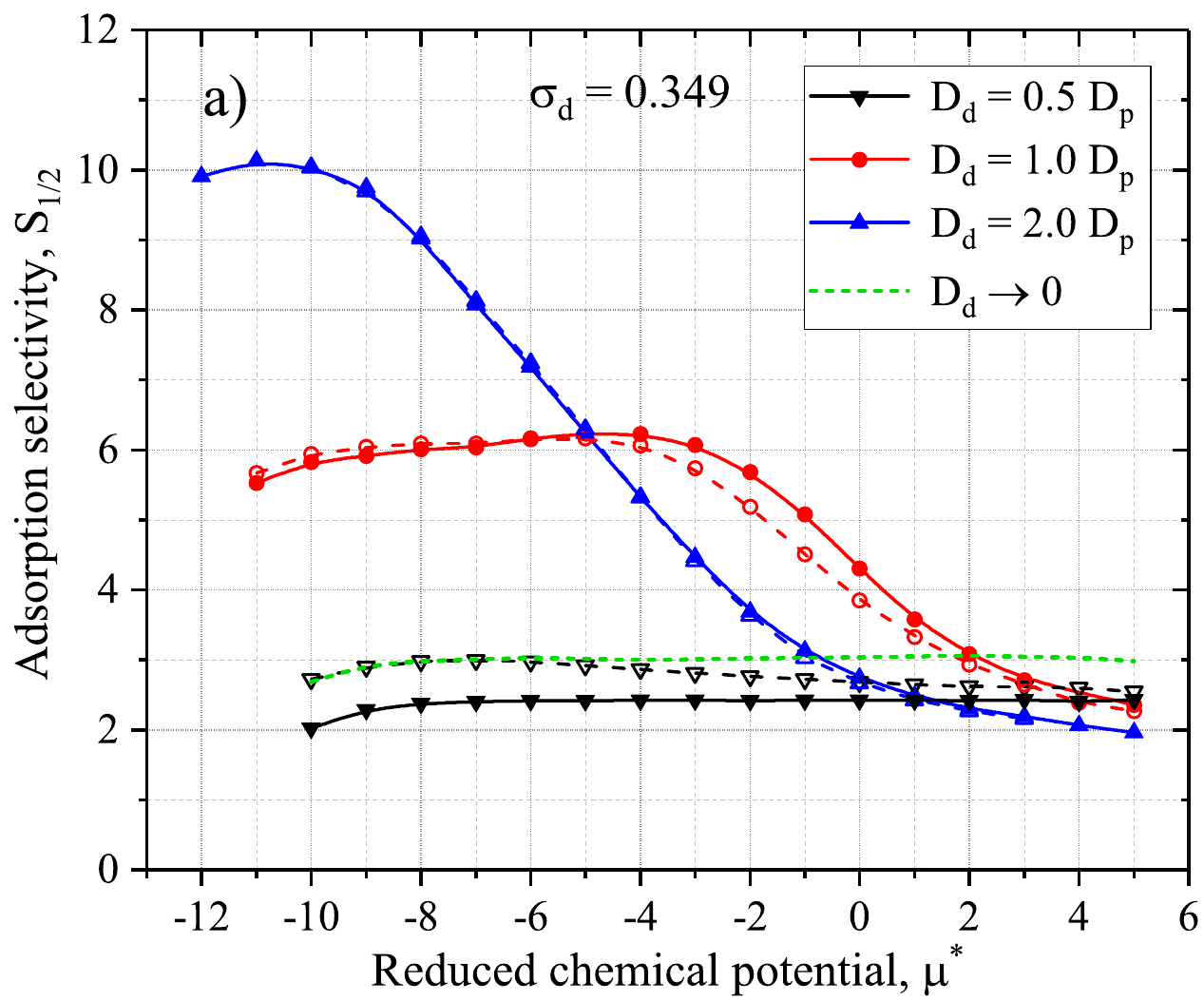}
	\includegraphics[width=0.47\linewidth]{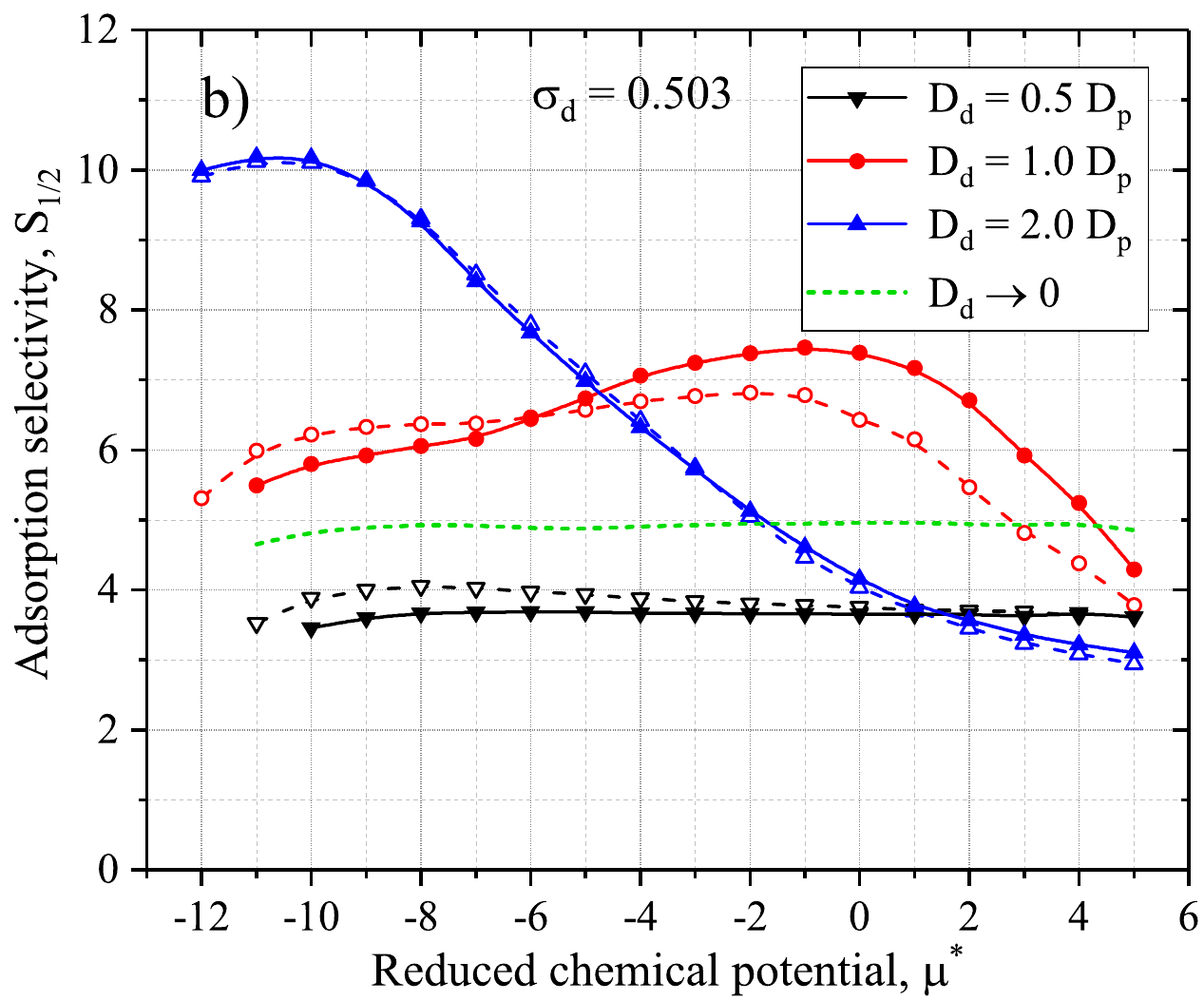}	
	\caption{(Colour online) Adsorption selectivity $S_{1/2}=\sigma_1/\sigma_2$ for a binary mixture of disk-like particles on a patterned
		adhesive surface with domain surface coverage
		$\sigma_{\rm d}=0.349$~(panel~a) and $\sigma_{\rm d}=0.503$~(panel~b).
        The line colours and styles correspond to the same parameters as in figure~\ref{fig:isotherms}.
	}
	\label{fig:selectivity}
\end{figure}

Based on the adsorption isotherms shown in figure~\ref{fig:isotherms}, the
adsorption selectivity was calculated as
$S_{1/2}=\sigma_1/\sigma_2$. The resulting dependence of $S_{1/2}$ on the 
chemical potential is shown in figure~\ref{fig:selectivity}. In all cases,
$S_{1/2}>1$, which confirms the preferential adsorption of species `1', as expected
from its stronger attraction to the adhesive domains. However, the magnitude of
the selectivity is not determined by the ratio of the attraction strengths alone.
It also strongly depends on the domain size, the domain surface coverage, and the
chemical potential.
For $\sigma_{\rm d}=0.349$ (figure~\ref{fig:selectivity}a), the weakest
selectivity is observed for the smallest domains,
$D_{\rm d}/D_{\rm p}=0.5$. In this case, the adhesive regions are numerous and
spatially distributed over the surface, so that both species experience a more
averaged attractive field. As a result, the selectivity remains relatively low and
varies just weakly with $\mu^*$. A much stronger selectivity is obtained for
$D_{\rm d}/D_{\rm p}=1.0$, where the domain size is comparable to the particle
size. In this case, adsorption sites with a large particle--domain overlap are
preferentially occupied by the more strongly attracted particles of species `1',
leading to a clear enhancement of $S_{1/2}$ at low and intermediate chemical
potentials.

For larger domains, $D_{\rm d}/D_{\rm p}=2.0$, the selectivity is especially high
at low chemical potentials. This reflects the fact that, under dilute adsorption
conditions, the few particles present at the surface are mainly those of the more
strongly attracted species. However, with increasing $\mu^*$, particles of species
`2' also start to populate the surface, and the selectivity decreases. At high chemical potentials, 
the curves corresponding to different domain sizes
approach one another, indicating that the selectivity becomes less sensitive to
the domain geometry as the adsorbed layer becomes increasingly crowded.

A higher domain surface coverage, $\sigma_{\rm d}=0.503$
(figure~\ref{fig:selectivity}b), leads to a noticeable change in the dependence of
selectivity on the reduced chemical potential. For $D_{\rm d}/D_{\rm p}=0.5$, the selectivity remains almost constant over the
whole range of $\mu^*$, similarly to adsorption on an effectively uniform
attractive surface. For $D_{\rm d}/D_{\rm p}=1.0$, the
selectivity is substantially enhanced and reaches its largest values at
intermediate chemical potentials. This indicates that particle-sized domains are
particularly efficient in discriminating between the two species when favourable
adsorption sites are sufficiently available but not yet saturated. By contrast, for
$D_{\rm d}/D_{\rm p}=2.0$, the selectivity is high at low $\mu^*$ but decreases
monotonously as the surface becomes more populated.

The comparison with the limiting case $D_{\rm d}\to 0$ (figure~\ref{fig:selectivity}) further shows that an
effectively uniform attractive surface gives a smoother dependence of $S_{1/2}$ on
$\mu^*$, without features associated with finite-size surface domains.
It is seen that the finite-domain systems deviate substantially from this
reference behaviour, especially for $D_{\rm d}/D_{\rm p}=1.0$ and $2.0$. This
additionally demonstrates that selective adsorption can be controlled by the size of the adhesive domains 
and their capability to accommodate particles of different affinity.
Interestingly, the selectivity in the limiting case $D_{\rm d}\to 0$ is higher
than that for finite domains with $D_{\rm d}/D_{\rm p}=0.5$. This suggests that,
once the domains are already smaller than the particle size, a further reduction
of their size may slightly enhance selectivity by making the attractive field more
uniformly distributed over the surface.

{For the domain surface coverage $\sigma_{\rm d}=0.349$
(figure~\ref{fig:selectivity}a), the comparison between ordered and disordered
domain arrangements shows that positional disorder affects the selectivity mainly
at low and intermediate chemical potentials. This is the regime in which
adsorption is most sensitive to the local distribution of adsorption sites. At
higher $\mu^*$, the difference between ordered and disordered domains becomes
weaker, because particle crowding at the surface increasingly masks the details of
the domain pattern. In general, the effect of domain disorder is moderate -- it is most visible
for $D_{\rm d}/D_{\rm p}=0.5$ and $1.0$, whereas for
$D_{\rm d}/D_{\rm p}=2.0$ it remains rather small. At the higher domain surface
coverage, $\sigma_{\rm d}=0.503$ (figure~\ref{fig:selectivity}b), the effect of
domain disorder becomes more pronounced for $D_{\rm d}/D_{\rm p}=0.5$ and $1.0$, but is
still weak for $D_{\rm d}/D_{\rm p}=2.0$. Notably, domain disorder tends to enhance the selectivity 
at lower values of the chemical potential, whereas at higher values it has the opposite effect and reduces the selectivity.}

{
To summarize, the obtained results indicate that adsorption selectivity can be enhanced either 
by increasing the domains surface coverage or by choosing an appropriate geometry of
the adhesive pattern. For surfaces structured by circular adhesive domains, the
most efficient selectivity is generally achieved when the domain size is comparable
to the particle size, although larger domains may be more favourable at low
particle densities. Small domains produce a weaker selectivity, but their
performance can be slightly improved by a disordered spatial arrangement.
}

\section{Conclusions}

We studied a selective adsorption in a binary hard-disk mixture on patterned
adhesive surfaces using grand canonical Monte Carlo simulations. The two particle
species had equal diameters and equal chemical potentials in the bulk reservoir,
but different attraction strengths to the adhesive domains. This setup allowed us
to focus on affinity-driven selectivity and its dependence on the geometry of the
surface pattern.

The results show that the geometry of the adhesive pattern has a pronounced effect
on adsorption selectivity. In addition to the affinity difference between the two
species, the selectivity depends on the domain size, domain surface coverage, and
spatial arrangement (ordered or disordered) of the domains. 
The strongest effects are observed at low and intermediate chemical potentials, 
where particles compete for favourable adsorption positions. 
Domains comparable to the particle size enhance selectivity by creating 
adsorption sites with a large particle--domain overlap, which are preferentially
occupied by the more strongly attracted species. 
Larger domains can also produce high selectivity at low chemical
potentials, although this effect weakens as the surface becomes more populated. 
It was also shown that in the case of domains a half smaller than particles, 
a further decrease of the domain size can also increase selectivity,
because the surface approaches an effective uniform attractive surface with
different effective affinities for the two components.

These findings suggest that, for particle-sorting applications, selectivity can be
improved not only by increasing the difference in particle--domain affinity, but
also by optimizing the geometry of the adhesive pattern, particularly by changing the adhesive domain size. 
Further extensions of the present model may consider binary mixtures of particles
with different sizes, as well as asymmetric bulk conditions in which the two
components have different chemical potentials. Such cases would make it possible
to examine how size effects, particle--domain affinity, and bulk composition
jointly determine the selective adsorption on patterned adhesive surfaces.

\section{Acknowledgements}

This work was supported by STCU (Ukraine) Grant No. 7115, with funding from the U.S. National Academy of Sciences (NAS) and the U.S. Office of Naval Research Global (ONRG). Computer time for the reported simulations 
was provided by the Interdisciplinary Center for Computer Simulations (Lviv), which is supported by the NRFU Grant No.~2023.05/0019.


\bibliographystyle{cmpj}
\bibliography{cmpjxampl}

%
%

\ukrainianpart
\title{Монте-Карло моделювання селективної адсорбції двокомпонентних сумішей твердих дисків на структурованих адгезивних поверхнях}
\author{Н.~Кукаркін\refaddr{label1}, Т.~Пацаган\refaddr{label1,label2}}
\addresses{
\addr{label1} Інститут фізики конденсованих систем НАН України імені І.~Р.~Юхновського, вул. Свєнціцького, 1, 79011 Львів, Україна
\addr{label2} Iнститут прикладної математики та фундаментальних наук, Нацiональний унiверситет ``Львiвська Полiтехнiка'', вул. С.~Бандери~12, 79013 Львiв, Україна
}
%
%
%

\makeukrtitle
\begin{abstract}
Селективну адсорбцію у двовимірній моделі двокомпонентної суміші твердих
дисків на структурованих адгезивних поверхнях досліджено методом Монте-Карло у
великому канонічному ансамблі. Частинки двох сортів мають однакові діаметри та
рівні хімічні потенціали в об’ємі, але відрізняються силою
притягання до адгезивних доменів. Таким чином, селективність, зумовлена
відмінністю в афінностях, відокремлена від ефектів, пов’язаних з асиметрією
розмірів частинок та різницею у хімічних потенціалах. Структура поверхні
задається розміром доменів, їх поверхневим покриттям, а також впорядкованим або
невпорядкованим розташуванням адгезивних доменів круглої форми. Результати показують,
що селективність адсорбції суттєво залежить від геометрії адгезивного патерну,
особливо при низьких і проміжних хімічних потенціалах. Домени розміром, близьким
до розміру частинок, підсилюють селективність завдяки утворенню адсорбційних
ділянок із великим перекриттям між частинкою і доменом, тоді як більші домени можуть
забезпечувати високу селективність при низьких хімічних потенціалах. У випадку
малих доменів подальше зменшення їх розміру також може підвищувати селективність,
оскільки система наближається до ефективно однорідної притягальної поверхні з
різними ефективними афінностями двох сортів частинок.
\keywords тверді диски, структурована поверхня, селективна адсорбції, моделювання методом Монте-Карло
\end{abstract}

\end{document}